\documentclass[prd,superscriptaddress,amsfonts,amssymb,amsmath,showpacs,twocolumn]{revtex4}
    \usepackage{epsfig}
    \usepackage{graphics}  
    \usepackage{color}
    \date{\today}

\begin{document}

\title{Rescuing the concept of swimming in curved spacetime}
\author{Rodrigo Andrade e Silva}
\email{rasilva@ift.unesp.br}
\affiliation{Instituto de F\'\i sica Te\'orica, Universidade 
Estadual Paulista,
Rua Dr.\ Bento Teobaldo Ferraz, 271 - Bl.\ II, 01140-070, 
S\~ao Paulo, SP, Brazil}
\author{George E.\ A.\ Matsas}
\email{matsas@ift.unesp.br}
\affiliation{Instituto de F\'\i sica Te\'orica, Universidade 
Estadual Paulista,
Rua Dr.\ Bento Teobaldo Ferraz, 271 - Bl.\ II, 01140-070, 
S\~ao Paulo, SP, Brazil}
\author{Daniel A.\ T.\ Vanzella}
\email{vanzella@ifsc.usp.br}
\affiliation{Instituto de F\'\i sica de S\~ao Carlos,
Universidade de S\~ao Paulo, Cx.\ Postal 369, 13560-970, 
S\~ao Carlos, SP, Brazil}

\begin{abstract}
It has been argued that an extended, quasi-rigid body evolving freely  in curved spacetime
can deviate from its natural trajectory by simply performing cyclic deformations. More interestingly,
in the limit of rapid cycles, the amount of deviation, per cycle, would depend on the sequence of 
deformations but not on how fast they are performed -- like the motion of a swimmer
at low Reynolds number.
Here, however, we show that the original analysis which supported this idea
is inappropriate to investigate the motion of extended bodies in the context of general 
relativity,
rendering its quantitative results invalid and casting
doubts on the reality of this {\it swimming} effect. 
We illustrate this by showing that the original analysis
leads to a non-zero  deviation even in a scenario where no swimming can possibly  occur. 
Notwithstanding, by applying a fully covariant, local formalism, we show that
swimming in curved spacetime is indeed possible and  that, in general,  its magnitude can be of the 
same order as (fortuitously) anticipated --
although it is highly suppressed in
the particular  scenario where it was originally investigated.
\end{abstract}


\maketitle

\section{Introduction}
\label{sec:intro}

More than a hundred years have passed and the picture of reality provided by special and general 
relativity can still surprise 
(and sometimes deceive)
our Newtonian intuition 
built upon notions of absolute space and time. In an analysis published in 
2003~\cite{Wisdom}, 
J.~Wisdom concluded that an extended, quasi-rigid body, free from external non-gravitational forces,
could propel itself through 
{\it curved} space by
performing cyclic internal motions. In the limit of rapid cycles, the 
amount of spatial translation (or, more properly, deviation from the trajectory it would follow if rigid) per cycle
would depend on the sequence of conformational changes of the body but 
not on how fast they are performed, similarly to the motion of a swimmer
at low Reynolds number -- hence the term {\it swimming effect}. 
This would be a legitimate curved-space effect, absent in Newtonian gravity, where this deviation 
 goes to zero in the limit of rapid cycles (as stated in Ref.~\cite{WisdomPT} in response to
Ref.~\cite{Landis}).

However, as novel and appealing as Wisdom's idea may be, here we show that his original analysis
is inappropriate to investigate the motion of extended bodies in curved spacetimes. 
We illustrate this by repeating his analysis in a scenario where no swimming can possibly occur
and show that it leads to a non-zero, fictitious effect, which renders his previous quantitative results invalid and 
casts serious doubts on the reality 
of 
the swimming effect. Notwithstanding, 
coming to the rescue of Wisdom's idea, we use a covariant, local 
approach developed in the 1970's by W.~G.~Dixon~\cite{DixonI,DixonII,DixonIII,DixonIV} to show that 
swimming in curved spacetime is indeed possible and argue that, in general, it can be 
of the same order of magnitude as
the (fictitious) effect obtained in Ref.~\cite{Wisdom}.

Our paper is organized as follows. In Sec.~\ref{sec:fict} we recall Wisdom's result for an articulated tripod evolving in
Schwarzschild spacetime and apply the same approach to a tripod
in de Sitter spacetime. In this latter context, we obtain a nonzero net displacement for the tripod which is very similar to
the one obtained in the former case. However, in Sec.~\ref{sec:msst}, after presenting a summary of Dixon's formalism, we show that
no swimming can possibly occur in maximally-symmetric spacetimes, contradicting the result of the previous section and thus proving Wisdom's approch inappropriate.
In Sec.~\ref{sec:rescuing} we revisit Wisdom's tripod in Schwarzschild spacetime using Dixon's formalism and show that,  although possible, 
swimming in this case is highly suppressed in comparison to Wisdom's result. Finally, in Sec.~\ref{sec:discuss} we make some final remarks, rescuing the
curved-spacetime swimming effect in the general context.


\section{Fictitious spacetime swimming?}
\label{sec:fict}

For concreteness sake, we shall  reuse Wisdom's symmetric tripod made of four point particles,
one with mass $m_0$ at the vertex of the tripod and three with mass $m_1$, one at the end of each leg.
Aligning the tripod
symmetry axis along the radial direction of
a spherically-symmetric object with mass $M$, and changing the length $l$ of its
legs (measured by static observers) and the angle $\alpha$ between each leg and the symmetry 
axis (measured in a local stationary Lorentz frame) according to the cycle
\begin{eqnarray}
(l,\alpha) & =&
(l_0,\alpha_0)  \to (l_0+\delta l,\alpha_0) \to (l_0+\delta l,\alpha_0+\delta \alpha)  
\nonumber
\\
& &
\to(l_0,\alpha_0+\delta \alpha) 
\to (l_0,\alpha_0),
\label{cycle} 
\end{eqnarray}
Wisdom calculated a radial deviation (with respect to the motion it would have if rigid), per cycle, to be given by
\begin{eqnarray}
\delta r_0 \approx -\frac{3m_0 m_1}{(m_0+3 m_1)^2} \left(\frac{l_0}{r_0}\right)^2 \frac{GM}{c^2 r_0}
\sin\alpha_0 \,\delta l \,\delta \alpha,
\label{drWisdom}
\end{eqnarray}
where $r_0$ is the radial coordinate of the mass $m_0$ ($G$ is Newton's constant and $c$ is the speed of light). The result above 
would supposedly  hold in
the limit of rapid cycles (short periods -- more on this, later), small velocities (in comparison to $c$), 
and $l_0/r_0, \, GM/(c^2r_0) , \, \delta l/l_0, \, \delta \alpha/\alpha_0\ll 1$. 

Now, let us consider this same tripod executing the same cycle but evolving
in de Sitter spacetime -- a maximally-symmetric spacetime with positive curvature. 
In order to avoid subtleties involving time-dependent metric components,
we cover (a portion of) the spacetime with static coordinates, in which the line element takes the form
\begin{equation}
ds^2 = -\left(1-\frac{r^2}{\kappa^2}\right)c^2 dt^2+\frac{dr^2}{\left(1-{r^2}/{\kappa^2}\right)}
+r^2 d\Omega^2,
\label{ds2}
\end{equation}
where  $\kappa$ is a constant related to the spacetime curvature $R= 12/\kappa^2$,
$r\in [0,\kappa)$  is the radial
coordinate with respect to some arbitrary spatial point, $t$ is the time-like coordinate, and
$d\Omega^2= d\theta^2 + (\sin\theta)^2 d\phi^2$ 
is the line element of the unit sphere.
We conveniently orient the 
tripod symmetry axis along the radial direction $\theta = 0$, in such a way that its legs have constant
$\phi$ coordinates (for instance, $\phi = 0, 2\pi/3, 4\pi/3$).

Following the  reasoning of Ref.~\cite{Wisdom}, the Lagrangian of the tripod would be the sum of the Lagrangians
of each particle, supplemented by the constraints which relate the  position of the particles in terms of
$l(t)$ and $\alpha(t)$:
\begin{eqnarray}
L(r_0,\dot{r}_0, t)&=&-m_0c^2\sqrt{\left(1-\frac{r_0^2}{\kappa^2}\right)-\frac{\dot{r}^2_0/c^2}
{\left(1-\frac{r_0^2}{\kappa^2}\right)}}\nonumber \\
& &
-3 m_1c^2\sqrt{\left(1-\frac{r_1^2}{\kappa^2}\right)-\frac{\dot{r}_1^2/c^2}{\left(1-\frac{r_1^2}{\kappa^2}\right)}
-\frac{r_1^2\dot{\theta}_1^2}{c^2}},
\nonumber \\
\label{L}
\end{eqnarray}
where dots represent derivative with respect to $t$. Note that the total Lagrangian is a function of 
$(r_0,\dot{r}_0)$, associated with the motion of the vertex of the tripod, and of the {\it assigned}
functions $l(t)$ and $\alpha(t)$, which relate the positions of the other three particles, at $(r_1, \theta_1,\phi=0)$,
$(r_1, \theta_1,\phi=2\pi/3)$, $(r_1, \theta_1,\phi=4\pi/3)$,
with $r_0$. Going through all the steps which
led Wisdom to Eq.~(\ref{drWisdom}), we obtain, for the same cycle~(\ref{cycle}),
\begin{eqnarray}
\delta r_0 \approx \frac{3m_0 m_1}{(m_0+3 m_1)^2} \left(\frac{l_0}{\kappa}\right)^2
\sin\alpha_0 \,\delta l \,\delta \alpha.
\label{drdeSitter}
\end{eqnarray}
This result is, both qualitatively and quantitatively, very similar to Eq.~(\ref{drWisdom}), with the scale of
curvature of Schwarzschild spacetime, $GM/(c^2r^3)$, replaced by the scale of curvature of de Sitter
spacetime, $1/\kappa^2$ (the change in sign is due to the fact that ``free fall'' in de Sitter is towards
increasing values of $r$). In fact, the results are so similar and consistent with each other that it is difficult
to raise any objection to one of them without compromising the other. But, as we argue below, 
the  deviation obtained in Eq.~(\ref{drdeSitter}) is entirely fictitious.

\section{No-swimming in maximally-symmetric spacetimes}
\label{sec:msst}

The motion of extended bodies in general relativity was analyzed in detail, in a manifestly covariant and local way,  in a seminal series of 
papers by W.~G.~Dixon in the 1970's~\cite{DixonI,DixonII,DixonIII,DixonIV}.  
One great advantage of Dixon's formalism, in addition to being local and covariant, 
is that covariant {\it conservation} of the stress-energy-momentum tensor of the body -- which strictly implements the
``absence of external non-gravitational forces'' in curved spacetime -- can be enforced from the 
beginning~\cite{Note1}.
Here we shall collect only the results which are relevant to our covariant analysis of the swimming effect,
focusing  on their physical meaning and implications rather than on  technical definitions,
and refer the reader to the original papers for further details. (Some technical remarks whose omission we may find
unbearable
we include as endnotes.)

Let $T^{ab}$ be the stress-energy-momentum tensor of the body under consideration. Its ten independent
components describe, at each point,
the energy and momentum densities of the body, as well as local {\it internal} forces exchanged among its parts.  We assume, from the beginning, that $T^{ab}$ is covariantly conserved:
$\nabla_a T^{ab}=0$, where $\nabla_a$ stands for the covariant derivative compatible with the
spacetime geometry. Thus, under  reasonable
assumptions about $T^{ab}$~\cite{Note2} 
and somewhat weak conditions on the ``strength'' of the gravitational field (see
Sec.~2 of Ref.~\cite{Bei}), the following general results hold:
\begin{itemize}
\item[(i)] At each and every spacetime point $x$ in the world-tube of the body (i.e., at which $T^{ab}(x) \neq 0$), there is a  {\it unique} time-like, future-pointing, unit vector $n^a=n^a(x)$ 
such that the {\it total} four-momentum $p^a = p^{a}(x,n)$ of the body (properly defined with respect to $x$ and $n^a$; see Ref.~\cite{DixonI}) is entirely in the
direction of $n^a$. In essence, this means that there is a family of observers (those with four-velocity $u^a= c\, n^a$) 
according to whom the total {\it spatial} momentum of the body is zero;
\item[(ii)] There is a {\it unique}, covariantly defined, time-like curve $z(\tau)$, contained in the (convex hull of the) world-tube of the body, which can be consistently
identified as the world-line of its  {\it center of mass} (here $\tau$ is the proper-time along this world-line). It is worth mentioning that the tangent vector to this curve,  
$v^a(\tau)$, is not, in general, parallel to
the four-velocity $u^a = c\, n^a$ -- see (i) -- at $z(\tau)$. In other words, the center of mass of the body is {\it not} necessarily at rest
with respect to the observers who attribute zero spatial momentum to the body~\cite{Note3};
\item[(iii)] With respect to  $z(\tau)$ and $n^a(\tau)$ ($\equiv n^a$ at $z(\tau)$), the total angular momentum of the body (also properly defined in Ref.~\cite{DixonI}) is characterized by a
spin
four-vector $S^a(\tau)$ orthogonal to $p^a(\tau) \equiv p^a(z(\tau),n(\tau))$;
\item[(iv)] Along $z(\tau)$, the four-momentum $p^a(\tau)$ and the spin $S^a(\tau)$ change according to the coupled equations
\begin{eqnarray}
\frac{D p^a}{d\tau} &=&\frac{1}{2} S^{bc}v^dR_{bcd}^{\;\;\;\;\;a} + F^a,
\label{dp}
\\
\frac{D S^{ab}}{d\tau}&=&(p^a v^b-v^a p^b)+G^{ab},
\label{dS}
\end{eqnarray}
where $D/d\tau := v^b\nabla_b$ is the covariant derivative operator along $z(\tau)$, $S^{ab} = \epsilon^{abcd}S_c n_d$ ($\epsilon_{abcd}$ is the totally antisymmetric, Levi-Civita pseudo tensor),
and $F^a$ and $G^{ab}$ are force-like~\cite{Note4} 
and torque-like terms which can be expanded in terms of 
couplings between the Riemann curvature tensor 
$R_{abc}^{\;\;\;\;\;d}$ (and its derivatives) and higher multipole moments
(quadrupole, octupole, and so on) of $T^{ab}$ (with respect to $z(\tau)$ and $n^a(\tau)$). For now, their exact forms are not important;
\item[(v)] For each (if there is any) symmetry of the spacetime  with generator  $\xi^a$ (which thus satisfies $\nabla_a\xi_b+\nabla_b\xi_a = 0$), there is an associated constraint between $F^a$ and
$G^{ab}$ given by
\begin{eqnarray}
F^a\xi_a+\frac{1}{2}G^{ab}\nabla_a\xi_b = 0.
\label{FGconst}
\end{eqnarray}
(Noting that $G^{ab}$ is antisymmetric.)
\end{itemize}
\noindent
These  (together with the precise definitions of $p^a$, $S^{ab}$, center of mass, $F^a$, and $G^{ab}$)  are the main results of the extensive analysis performed by Dixon.

One immediate consequence  concerns maximally-symmetric (Minkowski, de Sitter, and anti-de Sitter)  spacetimes. In such spaces,
the number of symmetries 
is enough to make $\xi_a$ and $\nabla_a\xi_b$ completely independent at each point. Therefore, the number of constraints
given in~(v) completely determines $F^a = 0$ and $G^{ab}= 0$. This, in turn, implies (see Ref.~\cite{DixonI}):  (a)~$p^a(\tau)$ is parallel to $v^a(\tau)$, with proportionality 
factor $m(\tau)$ being {\it constant} (interpreted as the rest mass of the system), (b)~the spin $S^a(\tau)$ is conserved, and, more importantly, 
(c)~$v^a(\tau)$ is parallel-transported along itself. In other words, {\it the center of mass of the body follows an exact geodesic of the spacetime}.
No swimming is possible in maximally-symmetric spacetimes, contradicting the result obtained in Eq.~(\ref{drdeSitter}).

This proves, beyond doubt, that the original analysis of Ref.~\cite{Wisdom}
is inappropriate to investigate the motion of extended bodies in curved spacetime, even in the regime where its approximations are valid. 
Considering that this kind of analysis stands as the  {\it sole} basis of the swimming effect  {\it to date}, this would seem to drown the hopes of any spacetime
swimmer. 

\section{Rescuing Wisdom's tripod in Schwarzschild spacetime}
\label{sec:rescuing}

Notwithstanding the negative general result  in maximally-symmetric spacetimes, 
let us take a closer look at Eqs.~(\ref{dp},\ref{dS}). As already mentioned, the force-like and torque-like terms, $F^a$ and $G^{ab}$, 
depend on the coupling between the geometry of the spacetime and the multipole moments of $T^{ab}$. Under the same assumptions made
in Ref.~\cite{Wisdom} concerning length scales of the body being much smaller than  length scales introduced by the curvature of the spacetime, the dominant contribution
to $F^a$ and $G^{ab}$ comes, in general, from the quadrupole moment of $T^{ab}$:
\begin{eqnarray}
F^a &=&- \frac{1}{6}J^{bcde}\nabla^aR_{bcde} \nonumber \\
& &+ \,\left(
\begin{tabular}{c}
higher-multipole,\\
higher-curvature-derivative\\
terms
\end{tabular}
\right),
\label{Fa}
\\
G^{ab} &=& \frac{4}{3}R^{[a}_{\;\;\,cde} J^{b]cde}
\nonumber \\
& &+ \,\left(
\begin{tabular}{c}
higher-multipole,\\
higher-curvature-derivative\\
terms
\end{tabular}
\right),
\label{Gab}
\end{eqnarray}
where $J^{abcd}$ is the {\it reduced} quadrupole moment of $T^{ab}$, defined in Refs.~\cite{DixonI,DixonIII}, calculated in the zero-spatial-momentum frame and with respect to the center of mass. 
(Brackets
stand for antisymmetrization over the enclosed indices: $A^{[ab]}:=(A^{ab}-A^{ba})/2$.)
The exact expression for $J^{abcd}$ in terms of $T^{ab}$ (and $z(\tau)$) does not concern us now. It suffices to mention a few  facts.
First,  it has the same index symmetries as
the Riemann curvature tensor $R^{abcd}$ (therefore, only $20$ independent components instead of the $60$ expected -- hence the term ``reduced'').
Second -- and here is where all the intricate definitions of Refs.~\cite{DixonI,DixonIII} pay off --, the covariant conservation of $T^{ab}$ imposes {\it no} (algebraic or differential) constraints on $J^{abcd}$
-- nor on any
of the higher reduced multipole moments, for that matter. 
Therefore, 
at least in principle, $J^{abcd}$ can be independently assigned as a function of the proper-time $\tau$, reflecting  conformational changes prescribed for the body in its own rest frame, 
and then  Eqs.~(\ref{dp},\ref{dS}) (ten components in total)
can be integrated to determine the  evolution of $p^a$, $S^a$, and  $v^a$ (ten components in total) -- and, thus, the world-line $z(\tau)$ itself.
Such is the power of Dixon's formalism.

Now, let us apply this machinery to Wisdom's tripod in Schwarzschild spacetime. Right from the start, the symmetry of the setup vanishes most of the components of a tensor
with the index symmetries of $J^{abcd}$, except (in an orthonormal basis 
$\{e_0^a,e_1^a,e_2^a,e_3^a\}$ aligned with the coordinates $\{t,r,\theta,\phi\}$, respectively): $J^{0101}$, $J^{0202}= J^{0303}$, $J^{0212}= J^{0313}$, $J^{1212}= J^{1313}$, and
$J^{2323}$. Combining this with the Riemann curvature tensor,
one finds that the quadrupole contribution to the torque in Eq.~(\ref{Gab}) 
vanishes: 
$R^{[a}_{\;\;\,cde} J^{b]cde} = 0$. This, in turn, adds complication to the analysis of Wisdom's tripod because, then,   in order to calculate $G^{ab}$, one must look at the next order, which couples the (reduced)
octupole moment $J^{abcde}$ of its stress-energy-momentum tensor (whose typical order of magnitude is $m c^2 l^3$, where 
$m $  is the mass of the tripod and $l$ its typical linear size) to the covariant derivative of the curvature tensor ($\sim G M/(c^2r^4)$, where
$r$ is the radial coordinate of the center of mass of the tripod).
In spite of its intricate expression -- see Eqs.~(13.8,13.9,A1.2,5.35)  of 
Ref.~\cite{DixonIII} --, in the end the symmetries of the system and of $J^{abcde}$ (see Eqs.~(5.33,5.36) of Ref.~\cite{DixonIII}) lead, to zeroth-order in $\dot{r}/c$~\cite{Note5}, to
\begin{eqnarray}
G^{01} \approx \frac{4GM}{c^2r^4}\,J^{22110},
\label{G01}
\end{eqnarray}
with all other components of $G^{ab}$ vanishing.
In addition, we note that the symmetry of the setup also imposes $S^a = 0 = S^{ab}$, which forces 
$v^a$ and $p^a$ to be slightly misaligned in order for the right-hand side of Eq.~(\ref{dS}) to vanish. To 
first order in {\it both} $GM/(c^2 r)$ and
$v^1/c = \dot{r}/c$~\cite{Note6},
\begin{eqnarray}
p^a\approx mv^a+ \frac{4GM}{c^3r^4} \,J^{22110}\,e_1^a.
\label{va}
\end{eqnarray}

Turning attention to the force-like term, Eq.~(\ref{Fa}), the symmetries of $J^{abcd}$ combined with the Riemann curvature tensor lead to
\begin{eqnarray}
F^1 = -\frac{4GM}{c^2 r^4}\left(J^{0101}-J^{0202}+J^{1212}-J^{2323}\right),
\label{F1}
\end{eqnarray}
with all other components of $F^a$ vanishing.
Therefore, substituting Eqs.~(\ref{va},\ref{F1}) and $S^{ab}=0$ into Eq.~(\ref{dp}), we have, to first order in both
$GM/(c^2r)$ and $\dot{r}/c$,
\begin{eqnarray}
\frac{D (mv^a)}{d\tau} \approx -\frac{4GM\Lambda }{c^2r^4}\,e_1^a,
\label{dmv}
\end{eqnarray}
where
$\Lambda = J^{0101}-J^{0202}+J^{1212}-J^{2323}+\dot{J}^{22110}/c$, which is of order $m c^2 l^2$.
This equation of motion tells us how much the tripod's center of mass can accelerate (in the nonrelativistic regime) with respect to exact
geodesic motion simply due to the fact that, being extended, it experiences an inhomogeneous gravitational 
field (more properly, inhomogeneous tidal effect). 
This is not swimming but one can use this estimate to obtain  a stringent upper bound on the order of magnitude of the 
swimming effect  for Wisdom's setup.
In particular, no variation in the quadrupole and octupole moments of $T^{ab}$, in the regime assumed in Ref.~\cite{Wisdom}, can change the $r^{-4}$ dependence of the 
right-hand side of Eq.~(\ref{dmv}) to a less-negative 
power.
Therefore, this simple order-of-magnitude estimate is enough to prove that 
Eq.~(\ref{drWisdom}) cannot hold true. 

The reduced quadrupole and octupole
moments of the tripod, needed to determine $\Lambda$, 
could, in principle, be given as a function of $\tau$~\cite{Note7}. 
But more realistically, these moments 
should be calculated  from the prescribed internal motion of the tripod. In other words, they must be given in terms of 
$l$ and $\alpha$; not only through their instantaneous values
$l(\tau)$ and $\alpha(\tau)$ but also dependent on the instantaneous values of 
$\dot{l}(\tau)$, $\dot{\alpha}(\tau)$, $\ddot{l}(\tau)$, and $\ddot{\alpha}(\tau)$. The $(\dot{l}, \dot{\alpha})$-dependence can come in due to possible energy flows inside the body 
as its conformation changes, while  $(\dot{l}, \dot{\alpha},\ddot{l}, \ddot{\alpha})$-dependence  comes in due to {\it internal} forces (stresses) necessary to accelerate its different parts with respect to
the center of mass. Recall that $T^{ab}$ does, indeed, contain all this information.

Comparing two slightly different prescriptions for the time dependence of $l$ and $\alpha$ (differing by $\delta l(\tau)$ and $\delta \alpha(\tau)$, respectively), 
we can expand the associated differences in the quadrupole and octupole moments
to first order in $\delta l$ and $\delta \alpha$ (and their derivatives) and, eventually, express the change in $\Lambda$ as
\begin{eqnarray}
\delta {\Lambda} = \sum_{q\in {\cal I}} \frac{\partial {\Lambda}}{\partial q}\,\delta q(\tau),
\label{deltaJtot}
\end{eqnarray}
where ${\cal I} = \{l,\alpha,\dot{l},\dot{\alpha},\ddot{l},\ddot{\alpha} \}$~\cite{Note8}. Then, 
from here on, the analysis follows as in Ref.~\cite{Wisdom}.
Using Eq.~(\ref{deltaJtot}) in Eq.~(\ref{dmv})
and working in the same regime as Ref.~\cite{Wisdom} [see below Eq.~(\ref{drWisdom})], the following approximate result holds in the limit of 
short periods of time~\cite{Note9}:
\begin{eqnarray}
m\delta r \approx  -\frac{4GM }{c^2r^4}\left(
 \frac{\partial {\Lambda}}{\partial \ddot{l}}\,\delta l+\frac{\partial {\Lambda}}{\partial \ddot{\alpha}}\,\delta \alpha\right).
\label{dreq}
\end{eqnarray}
Applying this result to the sequence of deformations~(\ref{cycle}),
a non-zero net 
deviation  in the radial position of the center of mass of the tripod, at the end of each cycle, 
only arises  if the right-hand side of Eq.~(\ref{dreq}) 
is {\it not} an exact differential in the parameter space $(l,\alpha)$. In that case, it is given by
\begin{eqnarray}
\delta r
\approx-\frac{4GM }{mc^2r^4} \left. \left(\frac{\partial^2 {\Lambda}}{\partial \alpha\partial \ddot{l}}-\frac{\partial^2 {\Lambda}}{\partial l \partial \ddot{\alpha}}\right)
\right|_{(l_0,\alpha_0)}\delta l \,\delta \alpha.
\label{drcycle}
\end{eqnarray}

In order to proceed beyond this point, we must calculate explicitly the relevant quadrupole and octupole 
components
which contribute to
$\Lambda$.
More precisely, we only need those quadrupole components which carry dependence on $\ddot{l}$ or $\ddot{\alpha}$
and the octupole components which depend on $\dot{l}$, $\dot{\alpha}$, $\ddot{l}$ or $\ddot{\alpha}$. 
Splitting the stress-energy-momentum tensor of the tripod as $T^{ab} = T^{ab}_{(m)}+T^{ab}_{(\ell)}$, with $T^{ab}_{(m)}$ and $T^{ab}_{(\ell)}$
separately
describing the point masses and the legs, respectively, only $T^{ab}_{(\ell)}$ shows explicit dependence 
on $\ddot{l}$ or $\ddot{\alpha}$.
This can be easily understood considering that the legs are  responsible for transmitting the {\it inner} forces which
ensure the prescribed changes in the relative
positions of the point masses.
On the other hand, modeling the legs as 
ideal rods (i.e., rods whose total masses and momenta can be neglected in the 
low-velocity regime
so that they transmit 
forces and torques 
integrally), one can verify that only $T^{ab}_{(m)}$ contribute to $\dot{J}^{22110}$ to leading order
-- due to the {\it momentum-density} distribution.
Hence, although the point masses of the tripod
dominate, through their mass distribution, the value of ${\Lambda}$ 
-- which determines how much the tripod deviate from  geodesic motion
for being extended --, it is the coupling of the spacetime curvature to (moments of) momenta and stresses 
 -- an inherent feature of relativistic gravity theories -- 
which impel the tripod to swim in this scenario.

For concreteness sake, 
considering only  leading-order curvature terms in expression~(\ref{drcycle}) and modeling the tripod's legs as ideal
rods (in the sense defined above), 
we can compute the relevant components of $J^{abcd}$ and $J^{abcde}$, 
which are $J_{(\ell)}^{1212}$, $J_{(\ell)}^{2323}$ and $J_{(m)}^{22110}$, as in flat space
(see Refs.~\cite{DixonI,DixonIII}):
\begin{eqnarray}
J_{(\ell)}^{ijij} &=& \frac{1}{4}\int d^3x\left[x^i x^i T_{(\ell)}^{jj}+x^jx^j T_{(\ell)}^{ii}-2 x^ix^j T_{(\ell)}^{ij}\right],\nonumber \\
\label{Jijij}
\\
J_{(m)}^{iijj0} &=& \frac{1}{8}\int d^3x\,x^i x^j\left[ x^i T_{(m)}^{j0}-x^j T_{(m)}^{i0}\right],
\label{Jiijj0}
\end{eqnarray}
where $\{x^i\}_{i=1,2,3}$ is a (local) Cartesian 
coordinate system whose origin is located at the center of mass of the tripod and with
$x^1$ aligned with the radial direction. The evaluation of $J_{(\ell)}^{1212}$ and $J_{(\ell)}^{2323}$ depends on
how the inner forces, leading to a given sequence of conformational changes, are exerted. In  
particularly simple realizations, in which each leg is subject only to forces in the plane containing the 
leg and the tripod's symmetry axis, one obtains $J_{(\ell)}^{2323}= 0$ and the contribution coming from
$J_{(\ell)}^{1212}$ and $J_{(m)}^{22110}$ leads
to
\begin{eqnarray}
\delta r \approx -
 \frac{GM}{c^2 r} \left(\frac{l_0}{r}\right)^3
 f(k,\alpha_0)\,\sin(2\alpha_0)\,
\delta l \,\delta \alpha,
\label{drfinal}
\end{eqnarray}
where $k :=3 m_1/(m_0+3 m_1)$ and $f(k,\alpha_0)$ is a dimensionless function whose exact
expression still depends on further details of how these forces are exerted on each 
leg~\cite{Note10}. 
This result
-- and, more generally, Eq.~(\ref{drcycle}) --
replaces the 
one given in Eq.~(\ref{drWisdom}) and it shows
that in the scenario
analyzed in Ref.~\cite{Wisdom} swimming is a much subtler effect. In fact, for a one-meter-long
tripod near  Earth's surface, swimming is suppressed by a $10^{-7}$ factor in comparison to
Eq.~(\ref{drWisdom}).

\section{Discussion: the curved-spacetime swimming effect}
\label{sec:discuss}

In Ref.~\cite{Wisdom}, Wisdom made use of a  
classical-mechanics analogue -- that of  an articulated, varying-length  
bipod moving on the surface of a 2-sphere without external 
tangential 
forces --
in order to 
ingeniously motivate the idea of swimming in curved spaces. There, the ``swimming'' is indeed of order
$(l/r)^2$, where $r$ is the radius of the 2-sphere. 
(This unfortunate similarity with Eq.~(\ref{drWisdom})
may  explain why the latter  stood undisputed for so long.)
It is interesting to point out
that Dixon's formalism (with some possibly minor changes to account for lower dimensionality) 
can be applied to this classical-mechanics system, accounting for the
same order-$(l/r)^2$  effect. But differently than what occurs for the
tripod in the scenario analyzed in Ref.~\cite{Wisdom}, 
it is the quadrupole contribution to the torque-like term $G^{i0} \propto R^{[i}_{\;\;\,cde} J^{0]cde}$ in Eq.~(\ref{dS}) 
which propels the center of mass of the
bipod along the 2-sphere. 
Due to symmetries of the body and of the spacetime (which can be seen as the $1+2$~Einstein static universe), 
this  torque-like term cannot induce 
rotation on the bipod; instead, it is cancelled by the ``misalignment'' term $p^i v^0-v^i p^0$.
As a result, at each cycle, the center of mass of the bipod is propelled along the 2-sphere 
($v^i \neq 0$)
while its total
spatial momentum is kept null (for $F^a = 0$ 
due to the constraints imposed by the geometric symmetries via Eq.~(\ref{FGconst})). 
Interestingly enough, a similar conclusion holds for a bipod in $1+3$ Einstein static universe.

The application of Dixon's formalism to the bipod on the 2-sphere  
serves more than simply as
a consistency check: it also shows that swimming in curved spacetimes can be of the same order of magnitude as
anticipated by Wisdom.
For instance, it is quite possible that Wisdom's tripod in Schwarzschild spacetime can swim at order 
$GMl^2\delta l\delta \alpha/(c^2r^3)$ in scenarios where $G^{ab}=R^{[a}_{\;\;\,cde} J^{b]cde}\neq 0$ in Eq.~(\ref{dS}) or 
$S^{ab}\neq 0$ in Eq.~(\ref{dp}) -- perhaps in orbital motion. Be it as it may,
one should use Dixon's covariant, local formalism, summarized in Eqs.~(\ref{dp},\ref{dS},\ref{Fa},\ref{Gab}),
in order not to arrive at fictitious effects and also to identify body shapes and internal cyclic motions  which might 
lead to optimal swimmers in 
a given spacetime.

\acknowledgments

The authors acknowledge full~(R.\ S.)~and partial~(G.~M.\ and D.\ V.)~financial support from 
S\~ao Paulo Research Foundation (FAPESP) under Grants No.~2015/10373-4,
2015/22482-2, and 2013/12165-4, respectively. G.\ M.\ also acknowledges partial financial support from
Conselho Nacional de Desenvolvimento Cient\'\i fico e Tecnol\'ogico (CNPq).

\end{document}